\documentclass[prb,preprint,showpacs,preprintnumbers,amsmath,amssymb]{revtex4}

\usepackage{graphicx}

\newcommand{\be}{\begin{equation}}
\newcommand{\ee}{\end{equation}}
\newcommand{\bea}{\begin{eqnarray}}
\newcommand{\eea}{\end{eqnarray}}
\newcommand{\eq}[1]{Eq.~(\ref{#1})}
\newcommand{\fig}[1]{Fig.~\ref{#1}}
\def\beq{\begin{eqnarray}}
\def\eeq{\end{eqnarray}}
\def\beqa{\begin{eqnarray}}

\def\eeqa{\end{eqnarray}}

\def\vq{{\bf q}}

\def\vk{{\bf k}}

\begin{document}

\title{Mean-field theory on a coupled system of ferromagnetism and electronic nematic order}

\author{Hiroyuki Yamase}
\affiliation{
Max-Planck-Institute for Solid State Research, D-70569 Stuttgart, Germany\\
and National Institute for Materials Science, Tsukuba 305-0047, Japan
}

\date{\today}

\begin{abstract}
We analyze an effective model on a square lattice 
with two types of forward scattering interactions, which,  respectively, 
drive ferromagnetism (FM) and electronic nematic order via 
a $d$-wave Pomeranchuk instability ($d$PI). 
The FM and $d$PI in general compete with each other and they 
are typically separated by a first order phase boundary 
in the plane of the chemical potential and temperature. 
Nevertheless there is a parameter region where 
the $d$PI occurs inside the FM phase, leading to their coexistence. 
We also study the effect of a magnetic field by choosing a chemical potential 
where the ground state is paramagnetic without a field. 
In this case, instead of FM, 
the $d$PI competes with a metamagnetic instability. The latter occurs above a threshold 
strength of the FM interaction and otherwise the $d$PI is stabilized with a dome-shaped phase 
diagram in the plane of a magnetic field and temperature. The FM interaction shifts 
the center of the dome to a lower field, accompanied by a substantial reduction of 
the field range where the $d$PI is stabilized and by an extension of the first order 
part of the transition line, although the maximal critical temperature does not change. 
The experimental phase diagram of 
the bilayer ruthenate Sr$_{3}$Ru$_{2}$O$_{7}$ can be well captured 
by the present theory. 
\end{abstract}

\maketitle

\section{Introduction}
In the nematic liquid crystal,\cite{deGennes93}  rodlike molecules have 
a preferred orientation. 
This state is characterized by breaking of orientational symmetry, 
retaining other symmetries of the system. 
Electronic analogues of the nematic liquid crystal attract much interest.  
Electrons have spin and the direction is defined in spin space. 
Using spin degrees of freedom, a {\it spin} nematic state is studied 
in quantum spin systems.\cite{andreev84,misguich05} 
Electrons also have orbital degrees of freedom. 
With orbital order such as an occupation difference between the $d_{yz}$- and $d_{zx}$-orbital 
in a $d$-electron system, electrons may break orientational symmetry  
without any additional symmetry breaking, leading to 
an {\it orbital} nematic state.\cite{raghu09,wclee09}  
Ferropnictides are possible materials for such a state.\cite{kasahara12,onari12} 
On the other hand, the orientation cannot be defined for charge itself. 
However, a nematic state can be realized by using a charge degree of freedom. 
Two routes toward a {\it charge} nematic state are proposed. 
When the system is close to a charge stripe order, 
namely one-dimensional-charge order, where both translational and orientational 
symmetry are broken, fluctuations of charge stripes may restore 
the former but the latter may be still broken.\cite{kivelson98}
The charge nematic order can be obtained also without invoking charge stripes. 
It was found theoretically that the two-dimensional $t$-$J$\cite{yamase00} 
and Hubbard\cite{metzner00} models exhibit a tendency toward 
a $d$-wave Pomeranchuk\cite{pomeranchuk59} instability ($d$PI). 
In this state, the Fermi surface expands along the $k_x$ direction 
and shrinks along the $k_y$ direction, or vice versa, whereas in a real space representation 
the nearest neighbor hopping is effectively enhanced 
along one direction and suppressed along the other direction. 

The $d$PI was extensively studied not only in the $t$-$J$\cite{yamase00,miyanaga06,edegger06,bejas12} and Hubbard\cite{metzner00,wegner02,neumayr03,okamoto10,su11,buenemann12} models, 
but also in phenomenological models,\cite{khavkine04,yamase05}  
a model with central forces,\cite{quintanilla06,quintanilla08}  
general Fermi liquid schemes,\cite{lamas08,zverev10}  and 
continuum (not lattice) models.\cite{oganesyan01,barci03,nilsson05,lawler06,zacharias09,maslov10} 
Mean-field theory of the $d$PI\cite{khavkine04,yamase05}  
showed that the $d$PI occurs around van Hove filling 
with a dome-shaped transition line. Typically the transition is second order at high temperature 
and changes to first order at lower temperature. 
The end points of the second order line are tricritical points. 
The mean-field phase diagram is characterized by a single energy scale, 
similar to the BCS theory of superconductivity, and thus various universal numbers 
were found.\cite{yamase05} 

Fluctuations of the $d$PI 
suppress the first order transition obtained in mean-field theory and when 
they are strong enough, the transition changes 
to be continuous even at zero temperature, leading to 
a quantum critical point.\cite{jakubczyk09,yamase11a}  
At the quantum critical point, $d$PI fluctuations lead to a 
non-Fermi liquid ground state.\cite{metzner03,dellanna06} 
At finite temperatures close to the $d$PI, thermal fluctuations become dominant.  
They turned out to truncate the original Fermi surface, 
leading to a Fermi-arc-like feature.\cite{yamase12} 

Signatures of nematicity were observed in cuprate superconductors. 
Neutron scattering measurements revealed a strong anisotropy of magnetic 
excitations in momentum space.\cite{hinkov04,hinkov07,hinkov08} 
The anisotropy showed strong temperature and doping dependences, 
which are well captured in terms of the competition of the $d$PI and 
singlet pairing formation.\cite{yamase06,yamase09}  
Transport measurements also revealed a very strong anisotropy of the 
Nernst coefficient,\cite{daou10} 
which was interpreted as a signature of charge nematic order.\cite{hackl09}  

There is growing evidence that 
the bilayer ruthenate Sr$_3$Ru$_2$O$_7$ (Sr327) 
exhibits a $d$PI in a strong magnetic field.\cite{grigera04,borzi07,rost09} 
In fact, many features observed in experiments were well understood 
in terms of the $d$PI, for example, the metamagnetic transition,\cite{kee05} 
the enhancement of the residual resistivity,\cite{doh07}  
the bilayer effect,\cite{puetter07,yamase09ab} 
the suppression of the critical temperature by impurities,\cite{ho08} 
and the spin-orbit effect.\cite{fischer10}  
Furthermore, the experimental phase diagram is very similar to that obtained in 
mean-field theory.\cite{yamase0710}  
In particular, it was found that the mean-field phase diagram 
is characterized by a single energy scale even in the presence of 
a magnetic field.\cite{yamase07c}  
Therefore there exist various universal 
ratios for a given chemical potential, which can be compared directly with experimental data. 
Although several universal ratios agree with the experimental data, ratios of 
the characteristic temperature and field  give one order of magnitude smaller 
than the experimental ones.\cite{yamase07c}  

This apparent inconsistency cannot be resolved by invoking different choices of parameters. 
The key may lie in the set of experimental indications that 
Sr327 is located close to a ferromagnetic instability: 
a large Wilson ratio,\cite{ikeda00} 
a uniaxial-pressure-induced ferromagnetic transition,\cite{ikeda04} and 
the presence of ferromagnetic fluctuations observed by the inelastic 
neutron scattering,\cite{capogna03} 
the nuclear spin-lattice relaxation rate,\cite{kitagawa05} 
and thermal expansion measurements.\cite{gegenwart06} 
Moreover several band calculations\cite{hase97,singh01} for Sr327 (without a field) 
suggested that the system is close to ferromagnetism (FM). 
Hence the presence of a ferromagnetic interaction is quite plausible in Sr327. 
In fact, early theoretical work\cite{millis02,binz04} for Sr327 
focused on the role of ferromagnetic interactions, especially in the context of a 
metamagnetic transition observed in experiments.\cite{perry01}

In this paper, we develop a mean-field theory by taking two types of 
forward scattering interactions, 
which drive the $d$PI and FM, respectively, into account. 
In the context of Sr327, it is interesting 
to explore how the mean-field phase diagram of the $d$PI obtained previously 
is modified by the presence of a ferromagnetic interaction and how well 
the experimental phase diagram of  Sr327 is captured. 
Furthermore, the interplay of the $d$PI and FM is interesting in its own right. 
While FM is an instability in the spin channel whereas the $d$PI is 
in the charge channel, both are instabilities in the particle-hole channel 
of $\vq$={\boldmath{$0$}} 
and do not break translational symmetry. 
Several theoretical analyses of microscopic models\cite{valenzuela01,fischer10,fischer11} 
actually suggested the presence of 
a ferromagnetic instability, which competes with the $d$PI. 
Therefore in a more general setting we study the interplay of 
the $d$PI and FM, and clarify possible scenarios in such a coupled system. 

We propose an effective model, suitable to address the interplay of the $d$PI and FM,  
and derive resulting phase diagrams. In Sec.~II, we introduce a forward scattering 
model and present results in Sec.~III by separating two cases: i) zero magnetic field ($h=0$) 
and ii) finite magnetic field ($h\ne0$). The latter case is relevant to Sr327. 
Conclusions follow in Sec.~IV.

\section{Model}
To analyze a coupled system of the $d$PI and FM, we consider the following 
Hamiltonian on a square lattice, 
\be
\mathcal{H}=\mathcal{H}_0+\mathcal{H}_{\phi}+\mathcal{H}_m+\mathcal{H}_Z\,.
\label{H}
\ee
The first term $\mathcal{H}_0$ is the kinetic term,  
\be
\mathcal{H}_0=\sum_{\vk \sigma} (\epsilon^{0}_{\vk} -\mu) c^{\dagger}_{\vk \sigma} c_{\vk \sigma} \,,
\label{H0}
\ee
where $c^{\dagger}_{\vk \sigma}$ $(c_{\vk \sigma})$ is a creation (annihilation) operator of 
an electron with spin $\sigma$ and momentum $\vk$; $\mu$ is the chemical potential. 
The electron dispersion is given by 
\be
\epsilon^{0}_{\vk}=-2 t (\cos k_x + \cos k_y) - 4 t' \cos k_x  \cos k_y
\ee
with $t$ and $t'$ being the nearest and second nearest neighbor hopping 
amplitudes, respectively.  

The second term $\mathcal{H}_{\phi}$ is a forward scattering interaction driving a $d$PI, 
\be
\mathcal{H}_{\phi}=-\frac{g_{\phi}}{2N}\sum_{\vk \vk' \sigma \sigma'} d_{\vk}d_{\vk'} c_{\vk \sigma}^{\dagger} c_{\vk \sigma} c_{\vk' \sigma'}^{\dagger} c_{\vk' \sigma'} \,,
\label{Hf}
\ee
where the coupling constant $g_{\phi}$ is positive, $d_{\vk}$ is a $d$-wave form factor 
such as $d_{\vk}=\cos k_x - \cos k_y$, and $N$ is the total number of lattice sites. 
This term describes the $d$-wave weighted density-density interaction 
with zero momentum transfer, which was  
obtained in microscopic models 
such as the $t$-$J$\cite{yamase00} and Hubbard\cite{metzner00,valenzuela01} models. 

The third term $\mathcal{H}_m$ describes an Ising ferromagnetic interaction,  
\be
\mathcal{H}_m=-\frac{g_m}{2N}\sum_{\substack{\vk \vk' \\ \sigma \sigma'}} \left(c_{\vk \sigma}^{\dagger}\frac{\sigma}{2} c_{\vk \sigma} \right) 
\left(c_{\vk' \sigma'}^{\dagger}\frac{\sigma'}{2} c_{\vk' \sigma'} \right) \,,
\label{Hm}
\ee
where $g_m (> 0)$ is a coupling constant and 
$\sigma=+1$ and $-1$ for up-spin and down-spin, respectively. 
This interaction is obtained by focusing on the spin-spin interaction with a spin quantization 
axis parallel to the $z$ direction and by extracting 
a scattering process with zero momentum transfer. 
Therefore the interaction described by $\mathcal{H}_m$ is 
appropriate when the system has a strong spin anisotropy 
as well as dominant forward scattering processes of electrons. 
The interaction of $\mathcal{H}_m$ is also obtained by considering 
a mean-field analysis of spin rotational invariant interactions. 
For instance,  in the case of the Hubbard onsite interaction 
$U\sum_{i} n_{i \uparrow}n_{i \downarrow}$, our coupling constant is given by $g_m=2U$.

The last term $\mathcal{H}_Z$ is the Zeeman energy, 
\be
\mathcal{H}_Z=-\frac{h}{2}\sum_{\vk \sigma} \sigma c_{\vk \sigma}^{\dagger} c_{\vk \sigma}\,.
\ee
Here $h$ is an effective magnetic field given by $h=\mathfrak{g}\mu_{B}H$, 
with $\mathfrak{g}$ being a $g$ factor, 
$\mu_B$ is the Bohr magneton, and $H$ is a magnetic field. 

The terms of $\mathcal{H}_{\phi}$ and $\mathcal{H}_m$ describe pure forward scattering 
interactions of electrons. Thus fluctuations around the mean-field 
vanish in the thermodynamic limit. 
In other words, mean-field theory solves our Hamiltonian exactly in the limit of 
$N \rightarrow \infty$. 

The order parameter of the $d$PI is defined by 
\be
\phi=\frac{g_{\phi}}{N} \sum_{\vk \sigma} d_{\vk} 
\langle c_{\vk \sigma}^{\dagger} c_{\vk \sigma} \rangle \,.
\label{phi}
\ee 
This quantity becomes finite only if the system breaks square lattice symmetry 
because of the presence of the $d$-wave form factor. 
FM order is defined by 
\be
m=\frac{g_m}{2N}\sum_{\vk \sigma} \sigma   
\langle c_{\vk \sigma}^{\dagger} c_{\vk \sigma} \rangle \,, 
\label{m}
\ee
where we include the coupling constant $g_m$ in the definition of $m$; 
while the magnetization is then given by $m/g_m$, we may refer to $m$ as 
magnetization, as long as no confusion occurs. 
We decouple the interaction terms (\ref{Hf}) and (\ref{Hm}) 
by introducing the order parameters $\phi$ and $m$, and obtain the mean-field 
Hamiltonian, 
\be
\mathcal{H}_{MF}=\sum_{\vk \sigma} \xi_{\vk \sigma} 
c_{\vk \sigma}^{\dagger} c_{\vk \sigma} + \frac{N}{2g_m} m^2  + \frac{N}{2g_{\phi}} \phi^2 \,,
\ee
where the renormalized dispersion is given by 
\be
\xi_{\vk \sigma}=\epsilon^{0}_{\vk} -\frac{\sigma}{2}(m+h) -d_{\vk} \phi - \mu\,.
\label{band}
\ee
The grand canonical potential per site at temperature $T$ is obtained as 
\be
\omega=-\frac{T}{N}\sum_{\vk \sigma} \log (1+ {\rm e}^{-\xi_{\vk \sigma}/T})  
+ \frac{1}{2g_m} m^2  + \frac{1}{2g_{\phi}} \phi^2 \,.
\label{omega}
\ee
The stationary condition of $\omega$ with respect to $\phi$ and $m$ leads to the 
self-consistent equations 
\bea
&&\phi=\frac{g_{\phi}}{N}\sum_{\vk \sigma} d_{\vk} f(\xi_{\vk \sigma})\,, 
\label{self-phi}\\
&&m=\frac{g_m}{2N}\sum_{\vk \sigma} \sigma f(\xi_{\vk \sigma})\,,
\label{self-m}
\eea
which we solve numerically. Here $f(\xi_{\vk \sigma})$ is the Fermi function.

\section{Results}
We fix $g_{\phi} /t =1$ throughout this paper unless otherwise noted 
and explore how the phase 
diagram of the $d$PI changes with increasing the FM interaction $g_m$. 
We first study the case of $h=0$ and then that of $h\ne0$.
As a band parameter, we choose $t'/t=0.35$, which was used for the study of 
Sr327.\cite{yamase0710,yamase07c} Since the presence of $t'$ turns out 
to play a crucial role to understand phase diagrams for $h=0$, 
we also study the case of $t'=0$ for $h=0$. 
Hereafter we set $t=1$ and all quantities with dimension of energy are in units of $t$. 

\subsection{Results for \boldmath{$h=0$}}
\subsubsection{Evolution of phase diagrams with increasing FM interaction} 

\begin{figure}[ht!]
\begin{center}
\includegraphics[width=6.5cm]{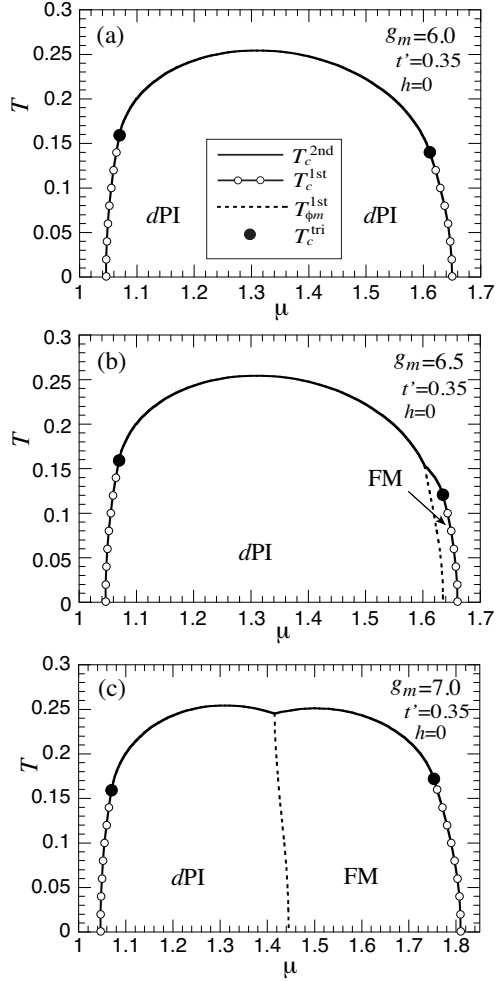}
\caption{Phase diagram in the $(\mu, T)$ plane for a sequence of couplings $g_m$. 
Transition from the paramagnetic to ordered phase is 
a second order $(T_c^{\rm 2nd})$ at high $T$ and a first order $(T_c^{\rm 1st})$ 
at low $T$; $T_c^{\rm tri}$ is the temperature at a tricritical point. 
A dashed line $(T^{\rm 1st}_{\phi m})$ denotes the first order phase boundary 
between the $d$PI and FM, which appears in (b) and (c). 
} 
\label{gm0-7h0}
\end{center}
\end{figure}

Figure~\ref{gm0-7h0} shows a sequence of phase diagrams for $g_m \leq 7.0$ 
in the plane of the chemical potential $\mu$ and temperature $T$. 
Because of the competition with the $d$PI, 
no FM instability occurs at least up to $g_m =6.0$ [\fig{gm0-7h0} (a)] 
and the phase diagram is occupied only by the $d$PI. 
As already clarified previously,\cite{khavkine04,yamase05} the $d$PI occurs 
below a dome-shaped transition line, with a maximal $T_c$ near the van Hove 
energy ($\mu_{\rm vH} =4t'=1.4$); a deviation from $\mu_{\rm vH}$ is due to 
the presence of $t'$, which breaks particle-hole symmetry. 
The phase transition is of second order at high 
temperature $(T_{c}^{\rm 2nd})$ 
and of first order at low temperature $(T_{c}^{\rm 1st})$. 
The end points of the 
second order line are tricritical points $(T_{c}^{\rm tri})$. 

For $g_m=6.5$, the FM interaction becomes strong enough to realize FM 
near the edge on the side of a high chemical potential [\fig{gm0-7h0} (b)]. 
The transition from the paramagnetic to FM phase is second order at high 
temperature, but the second order line ends at a tricritical point and 
changes to a first order line at low temperature. 
This feature is the same as the transition between the paramagnetic and $d$PI phase. 
The boundary of the $d$PI and FM is characterized by a first order transition $(T^{\rm 1st}_{\phi m})$. 

\begin{figure}[ht!]
\begin{center}
\includegraphics[width=6.0cm]{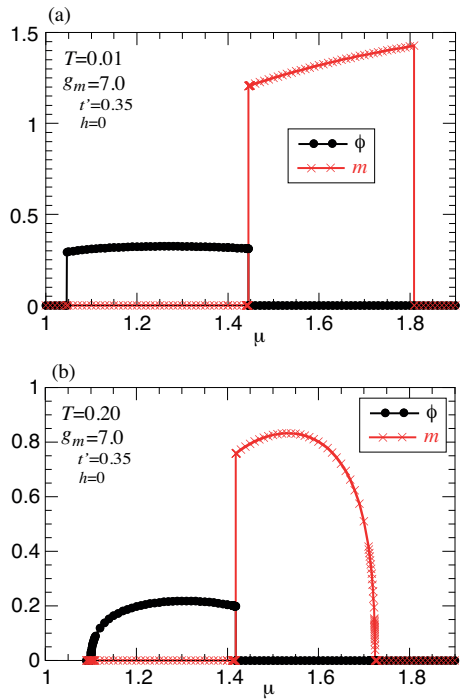}
\caption{(Color online) $\mu$ dependence of $\phi$ and $m$ at $T=0.01$ (a) 
and $0.20$ (b) for $g_m=7.0$. 
} 
\label{gm7h0order}
\end{center}
\end{figure}

As shown in \fig{gm0-7h0} (c), 
this first order phase boundary shifts to the middle of the phase diagram 
for $g_m=7.0$  and the FM becomes more stable.  
The order parameters are plotted as a function of $\mu$ in Figs.~\ref{gm7h0order} (a) and (b) 
at $T=0.01$ and $0.20$, respectively. At a low temperature ($T=0.01$), 
$\phi$ and $m$ show a jump 
at $\mu \approx 1.05$ and $1.81$, respectively, because of a first order transition from 
the paramagnetic phase. The $d$PI changes to the FM via a first order transition 
at $\mu \approx 1.45$ and there is no mixing of $\phi$ and $m$. 
At a high temperature ($T=0.20$), on the other hand, $\phi$ and $m$ develop 
continuously at $\mu \approx 1.10$ and $1.72$, respectively. 
The transition between the $d$PI and FM is however still of first order. 

\begin{figure*}[ht!]
\begin{center}
\includegraphics[width=10.5cm]{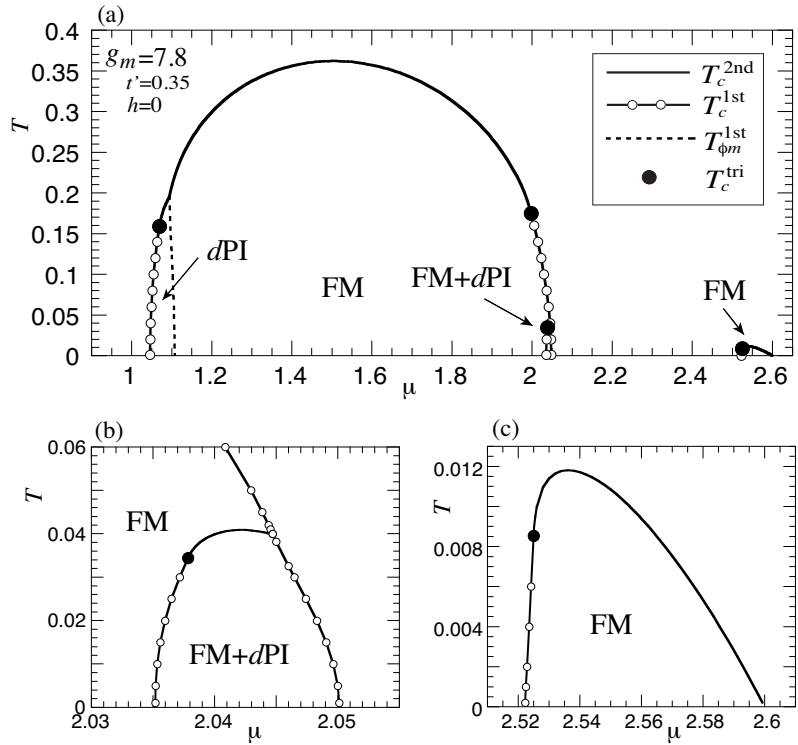}
\caption{Phase diagram in the $(\mu, T)$ plane for $g_m=7.8$.  
The regions near $\mu=2.04$ and $2.55$ are magnified in (b) and (c), respectively. 
} 
\label{gm7.8h0}
\end{center}
\end{figure*}

As expected, with further increasing $g_m$, the first order boundary 
between the $d$PI and FM shifts to a lower chemical potential. 
In fact, as shown in \fig{gm7.8h0} (a), the $d$PI is realized 
only near the edge of the dome for $g_m=7.8$. 
However, qualitative changes occur in the phase diagram. 
First, the coexistence of the $d$PI and FM is stabilized inside the FM phase 
near the edge of the first order line of the FM around $\mu=2.04$. 
This region is magnified in \fig{gm7.8h0} (b). 
The transition from the FM to the coexistence is first order at low temperature 
and becomes second order at high temperature. 
While one end point of the second order line at $\mu \approx 2.037$ 
is a tricritical point, the other end point at $\mu \approx 2.045$ is just a point touching 
the first order line of the FM. 
There is a direct first order transition from the paramagnetic phase to the coexistence
around $\mu = 2.05$. 
Second, an additional FM phase appears in $2.52 \lesssim \mu \leq 2.6$ as shown 
in Figs.~\ref{gm7.8h0} (a) and (c).  
This FM comes from the enhancement of the density of state at the band edge 
of $\mu=2.6$. A first order transition occurs only on the side of 
a lower chemical potential and the second order line disappears at the band edge. 
This band-edge FM is realized for $7.6 \lesssim g_m \lesssim 8.0$. 

\begin{figure}[t!]
\begin{center}
\includegraphics[width=7.0cm]{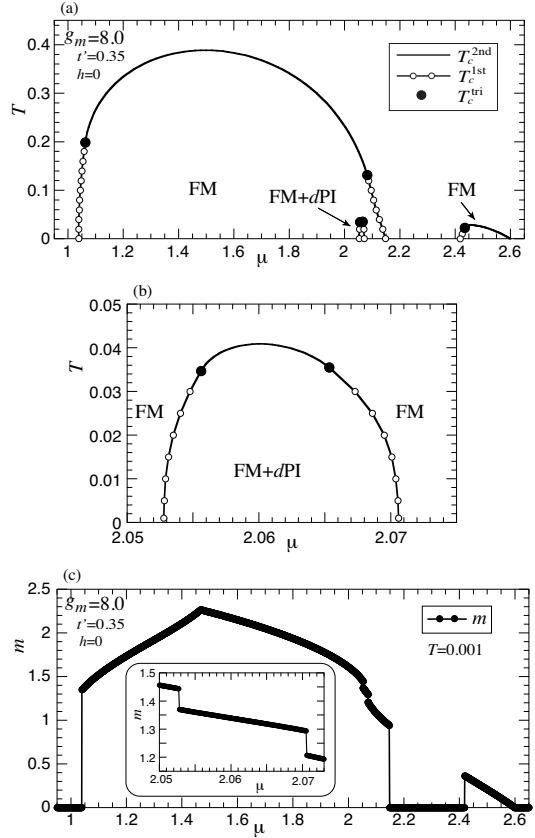}
\caption{(a) Phase diagram in the $(\mu, T)$ plane for $g_m=8.0$. 
The region of the coexistence around $\mu=2.06$ is magnified in (b). 
(c) $\mu$ dependence of $m$ at $T=0.001$. Two successive jumps 
around $\mu=2.06$ are magnified in the inset. 
} 
\label{gm8h0}
\end{center}
\end{figure}

For $g_m=8$, as shown in \fig{gm8h0} (a), the FM becomes dominant 
and a pure $d$PI phase is not stabilized. 
Instead the $d$PI is realized in coexistence with the FM around $\mu=2.06$, 
as magnified in \fig{gm8h0} (b).  
In contrast to the case of $g_m=7.8$ [\fig{gm7.8h0} (b)], the phase 
boundary of the coexistence is well separated from the first order line of the FM, 
leading to a phase diagram very similar to that of 
the pure $d$PI [\fig{gm0-7h0} (a)], but with a significant extension of the first 
order portion of the transition line; the reason for this will be explained later 
in terms of \eq{Landaua4}. 
The magnetization $m$ is plotted as a function of $\mu$ 
in \fig{gm8h0} (c) at low temperature. 
After the first order FM transition at $\mu \approx 1.05$, the value of $m$ increases with 
increasing $\mu$ and forms a cusp at $\mu \approx 1.45$ where the density of states of 
up-spin electrons is fully occupied and the system changes to a half-metallic state. 
For $\mu \gtrsim 1.45$, $m$ decreases since electrons with  down-spin increase 
whereas the electron density of up-spin remains unity. 
At $\mu \approx 2.05$ and $2.07$, $m$ exhibits a jump [see the inset of \fig{gm8h0} (c)] 
because of the presence of the 
coexistence of the $d$PI and FM, which occurs via a first order transition at low $T$. 
The magnetization $m$ vanishes discontinuously at $\mu \approx 2.15$, but appears 
again with a jump at $\mu \approx 2.42$ because of a first order transition associated 
with the band-edge FM. 
The magnitude of $m$ decreases monotonically and vanishes 
at the band edge of $\mu=2.6$.  
The system becomes a band insulator for $\mu > 2.6$.

\begin{figure*}[ht!]
\begin{center}
\includegraphics[width=11.5cm]{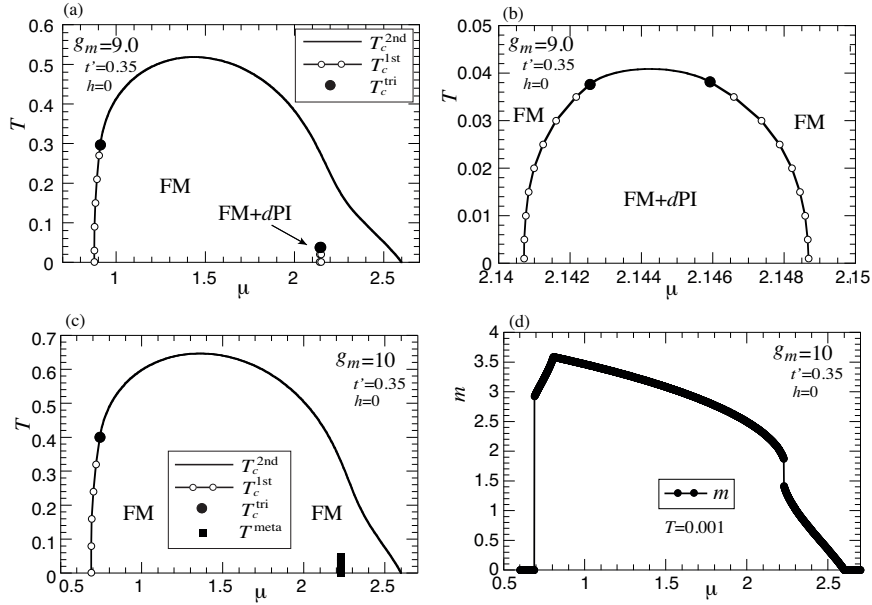}
\caption{(a) Phase diagram in the $(\mu, T)$ plane for $g_m=9.0$. 
The region of the coexistence is magnified in (b). 
(c)  Phase diagram for $g_m=10$. 
$T^{\rm meta}$ denotes the position where a metamagnetic transition occurs. 
(d) $\mu$ dependence of $m$ at $T=0.001$ for $g_m=10$. 
 } 
\label{gm10h0}
\end{center}
\end{figure*}

With further increasing $g_m$ (\fig{gm10h0}), 
the band-edge FM is absorbed into the main FM phase. 
A first order phase transition then occurs only on the lower side of $\mu$. 
Inside the FM, the coexistence of the $d$PI and FM is stabilized up to $g_m=9.8$. 
Figure~\ref{gm10h0} (a) is the representative phase diagram computed for 
$g_m=9$. In \fig{gm10h0} (b) the region of the coexistence of the $d$PI and FM is magnified. 
This phase diagram is very similar to that for $g_m=8$ [\fig{gm8h0} (b)] with the same 
maximal $T_c$, but with a further extension of the first order transition line. 
For $g_m \gtrsim 9.8$, however, the coexistence is replaced 
by a first order transition associated with a jump of the magnetization, 
namely a metamagnetic transition inside the FM, as denoted by solid squares 
in \fig{gm10h0} (c). 
The magnetization is potted as a function of $\mu$ at low $T$ in \fig{gm10h0} (d). 
The jump at $\mu \approx 2.23$ comes from the metamagnetic transition. 
The cusp at $\mu \approx 0.81$ indicates that the up-spin band is fully occupied 
in $\mu \gtrsim 0.81$, where the system becomes half-metallic.

\subsubsection{Discussions}
The coexistence of the $d$PI and FM is stabilized even for $g_m \gg g_{\phi}$ 
(Figs.~\ref{gm7.8h0}$-$\ref{gm10h0}).  
This is because of the presence of the van Hove singularity. 
After performing explicit calculations up to $g_m=10$, we confirm the 
van Hove singularity due to the down-spin band ($m>0$ is assumed) 
inside the FM phase for $g_m \gtrsim 7.8$. 
Around the van Hove filling, 
either the $d$PI or a metamagnetic transition occurs in our model, 
depending on energetics. 
We find that the coexistence of the $d$PI and FM is more favorable for 
$7.8 \lesssim g_m \lesssim 9.8$ and the metamagnetic transition for $g_m \gtrsim 9.8$. 

\begin{figure*}[th!]
\begin{center}
\includegraphics[width=11.5cm]{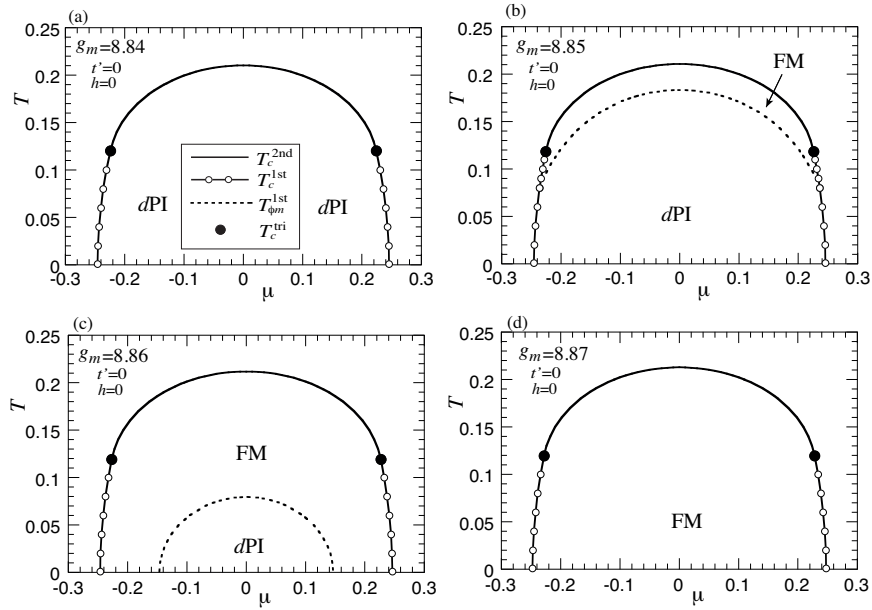}
\caption{Phase diagram in the $(\mu, T)$ plane for a sequence of 
couplings $g_m$ by setting $t'=0$.   
The phase diagram is occupied by the $d$PI in $g_m \leq 8.84$ (a) and by the FM 
in $g_m \geq 8.87$ (d). In a tiny range of $g_m$ [(b) and (c)], 
both FM and $d$PI are realized, but separated  
from each other by a first order boundary; the line of $T^{\rm 1st}_{\phi m}$ appears only in (b) and (c). 
} 
\label{tp0h0}
\end{center}
\end{figure*}

Our results shown in Figs.~\ref{gm0-7h0}$-$\ref{gm10h0} are very asymmetric with respect 
to the van Hove energy of the bare dispersion, which is given by $\mu_{\rm vH}=4t'=1.4$. 
This is because the presence of $t'$ breaks particle-hole symmetry. In fact,  for $t'=0$, 
the phase diagram becomes symmetric with respect to the axis of $\mu=0$. 
For $0 \leq g_m \leq 8.84$, the $d$PI is stabilized and no FM 
is realized [\fig{tp0h0} (a)]. For $g_m \gtrsim 8.85$, however, the $d$PI starts to be replaced 
by the FM phase from a higher temperature [\fig{tp0h0} (b)] and is stabilized only around 
$\mu=0$ at low $T$ for $g_m=8.86$ [\fig{tp0h0} (c)]. 
The $d$PI disappears already for $g_m=8.87$.
The change from the $d$PI  [\fig{tp0h0} (a)] to the FM phase [\fig{tp0h0} (d)] 
occurs in a very small range of $g_m$. 
In contrast to the case of Figs.~\ref{gm7.8h0}, \ref{gm8h0}, and \ref{gm10h0}, no coexistence 
of the $d$PI and FM is stabilized. Furthermore a band-edge FM does not appear.

Our results for $h=0$ are summarized as follows:  
i) in $0 \leq g_m \leq g_{m 1}$, only the $d$PI phase is realized, 
ii) in $g_{m 1} \leq g_m \leq g_{m 3}$, both $d$PI and FM are stabilized, 
but they are separated from each other by a first order transition line, 
iii) in $g_{m 2} \leq g_m \leq g_{m 4}$, the coexistence with $d$PI occurs inside the FM phase, 
and iv) in $g_{m 4} \leq g_m$, only the FM is realized. 
We have obtained $g_{m 1} \approx 6.5$, $g_{m 2} \approx 7.8$, $g_{m 3} \approx 7.9$, and 
$g_{m 4} \approx 9.8$ for $t'=0.35$,  leading to rich phase diagrams as shown in 
Figs.~\ref{gm0-7h0}, \ref{gm7.8h0}, \ref{gm8h0}, and \ref{gm10h0}. 
For $t'=0$, on the other hand, we have obtained 
$g_{m 1}\approx 8.84$, $g_{m 2}=g_{m 3}=g_{m 4} \approx 8.87$.  
As a result, a phase diagram is occupied by either the $d$PI or FM except for 
a tiny range of $g_m$.

\subsection{Results for \boldmath{$h\neq 0$}}
Next we examine the effect of a magnetic field, motivated by the experimental indication 
that Sr327 is paramagnetic in zero field and 
exhibits a nematic instability around 8 Tesla.\cite{grigera04,borzi07,rost09}    
Fixing the chemical potential $\mu=1$ and taking the field as a tuning parameter,  
we study how the phase diagram of the $d$PI 
evolves with increasing the ferromagnetic interaction.  

\begin{figure}[ht!]
\begin{center}
\includegraphics[width=6.5cm]{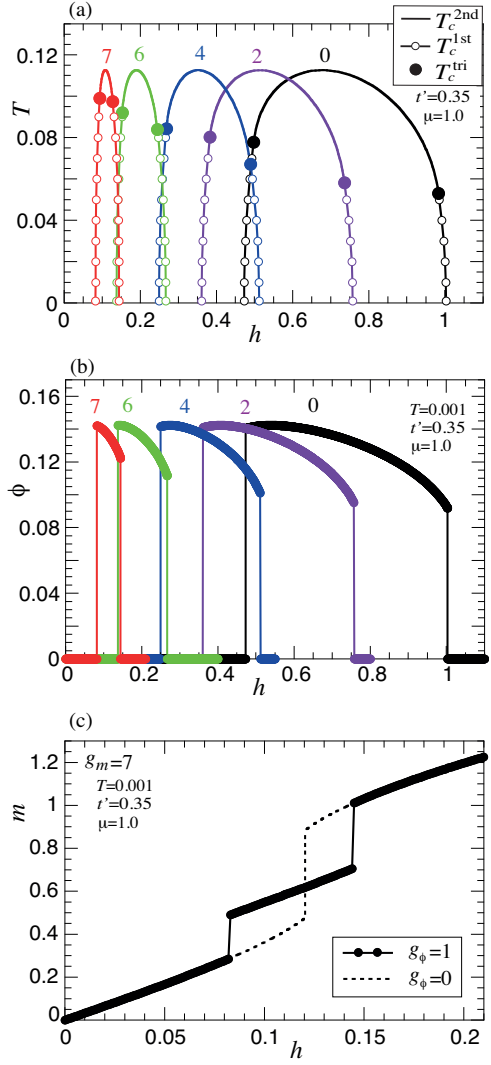}
\caption{(Color online) (a) Phase diagram in the $(h, T)$ plane for a sequence of 
couplings $g_m$; 
the value of $g_m$ is indicated near the maximal $T_c$. The $d$PI is stabilized 
inside the dome for each $g_m$. 
(b) $h$ dependence of $\phi$ at $T=0.001$ for a sequence of $g_m$. 
(c) $h$ dependence of $m$ for $g_m=7$ at $T=0.001$. The corresponding result for 
$g_{\phi}=0$ is also plotted. 
} 
\label{gm0-7h}
\end{center}
\end{figure}

Figure~\ref{gm0-7h} (a) is a set of phase diagrams of the $d$PI in the plane 
of a magnetic field and temperature for a sequence of $g_m$, 
showing four characteristic features: with increasing $g_m$, 
i) the $d$PI occurs in a lower field, 
ii) the field range where the $d$PI is stabilized shrinks substantially, 
iii) the first order part of the transition line extends and tricritical points are pushed up 
to higher temperatures, but 
iv) the maximal $T_c$ does not change. 

To understand these features, we consider a magnetic field $h_{\rm vH}$, 
at which the $\sigma$-spin band touches the van Hove energy, and the $d$PI 
is expected around that. 
From \eq{band}, $h_{\rm vH}$ fulfills for $\phi=0$ the relation, 
\be
\frac{\sigma (m+ h_{\rm vH})}{2} + \mu = \mu_{\rm vH}\,,
\label{vHh}
\ee
and the corresponding 
relation for the other spin band should be 
$-\sigma(m+h_{\rm vH})/2+\mu=2\mu-\mu_{\rm vH}$, where $\mu_{\rm vH}=4t'$. 
Since $\mu$ is fixed in our case, we obtain 
\be
h_{\rm vH} = 2 | \mu-\mu_{\rm vH}| -m\,. 
\label{vHh1}
\ee
While the magnetization is not fully linear in field in the entire field range we consider, 
we may invoke the equation obtained in linear response theory,  
\bea
&&m/g_m \approx \chi h_{\rm vH} \,,
\label{m-h} \\
&&\hspace{12mm} = \frac{\chi_0}{1-g_m \chi_0} h_{\rm vH}\,,
\eea
where $\chi$ is the full magnetic susceptibility, which is expressed by the non-interacting 
magnetic susceptibility $\chi_0$ as shown in the second line; 
the presence of $g_m$ on the left-hand side is due to our definition of $m$ 
[see \eq{m}].  We then obtain 
\be
h_{\rm vH}=2(1-g_m \chi_0) | \mu-\mu_{\rm vH}|\,,
\label{vHh2}
\ee
that it, the value of $h_{\rm vH}$ is reduced with increasing $g_m$. 
Since the $d$PI occurs around the 
van Hove energy, the $d$PI should be realized around a lower field with increasing $g_m$. 

Equation~(\ref{m-h}) is a rough approximation especially near a metamagnetic transition and 
the resulting \eq{vHh2} should be taken as qualitative understanding. 
To get quantitative understanding, we solve 
\eq{self-m} numerically under the condition of $\phi=0$ and \eq{vHh}. 
We then obtain $h_{\rm vH} \approx 0.61, 0.41, 0.22, 0.12$ for $g_m=2,4,6,7$, respectively; 
for $g_m=0$, on the other hand, $h_{\rm vH}=2|\mu_{\rm vH}-\mu|=0.8$ since $m=0$. 
The $d$PI indeed occurs around those fields in \fig{gm0-7h}. 

The range of a magnetic field where the $d$PI is stabilized becomes narrower for 
a larger $g_m$.  
As seen in \eq{band}, the sum of $m$ and $h$ plays a role 
as an effective field. 
Since $m$ becomes more susceptible to a field as $g_m$ becomes larger 
and furthermore $m$ is proportional to $g_m$ in our definition  [\eq{m}], 
the value of $h$ to stabilize the $d$PI is necessarily reduced.

The first order transition line extends with increasing $g_m$. 
To understand this, we expand the free energy \eq{omega} with respect 
to the order parameter of the $d$PI around $\phi=0$, 
\be
\omega(\phi; m) - \omega(0;m)
=\frac{1}{2}a_2 \phi^2 + \frac{1}{4!}a_4 \phi^4 + \cdots \,.
\label{Landau}
\ee
The coefficients of $a_2$ and $a_4$ are obtained as 
\bea
&& a_2= \frac{1}{g_{\phi}}\left(1+\frac{g_{\phi}}{N}\sum_{\vk \sigma} d_{\vk}^2 
f'(\xi^{0}_{\vk \sigma}) \right)\,,\\
\label{Landaua2}
&&a_4=\frac{1}{N}\sum_{\vk \sigma} d_{\vk}^{4} f'''(\xi^{0}_{\vk \sigma})
-3g_m\frac{\left(\frac{1}{2N} \sum_{\vk \sigma} \sigma d_{\vk}^2 f'' (\xi^{0}_{\vk \sigma})\right)^{2}}{1+\frac{g_m}{4N}\sum_{\vk \sigma} f' (\xi^{0}_{\vk \sigma})} \,,
\label{Landaua4}
\eea
where $\xi^{0}_{\vk \sigma}=\epsilon^{0}_{\vk} -\frac{\sigma(m+h)}{2}-\mu$ 
and $f', f'', f'''$ are the first, second, third derivative of the Fermi function.  
When $a_4$ becomes negative, a first order transition can occur.  
The second term on the right-hand side of \eq{Landaua4} 
originates from the $\phi$ dependence of $m$. 
The denominator of this term is positive close to the $d$PI and 
the numerator becomes in general finite when the spin symmetry is broken. 
Hence the second term is negative for $h\ne 0$. 
Furthermore the second term is proportional to $g_m$. 
Therefore the presence of the second term in \eq{Landaua4} 
leads to an extension of the first order transition line of the $d$PI 
and this effect becomes stronger for a larger $g_m$. 
The same argument explains the extension of the first order portion of the transition line 
in Figs.~\ref{gm8h0} (b) and \ref{gm10h0} (b), since the second term of 
\eq{Landaua4} becomes negative also in the FM phase. 

A second order transition is given by the condition $a_2=0$. 
Since $\mu$ is fixed, the quadratic term $a_2$ is a function of $\tilde{h}=m+h$. 
Suppose the maximal $T_c$ is obtained at $\tilde{h}_{\rm max}$, 
there can exist a field $h$ and a magnetization $m$, which 
give the same value of $\tilde{h}_{\rm max}$  for a different $g_m$, 
although the values of $m$ and $h$ themselves 
depend on $g_m$. This is actually the case up to $g_m=7.8$, leading to 
the same maximal $T_c$ in \fig{gm0-7h} (a). 
A similar consideration also explains the same maximal $T_c$ in Figs.~\ref{gm8h0} (b) 
and \ref{gm10h0} (b). Keeping in mind that our system is half-metallic in the range of 
$\mu$ where the coexistence is 
stabilized [see the discussion about \fig{gm8h0} (c)] 
and thus only the down-spin band is active, 
the coefficient $a_2$ becomes a function of the quantity 
$\tilde{\mu}=\frac{-m}{2}+\mu$ for $h=0$.  
We confirm the same value of $\tilde{\mu}$ at the maximal $T_c$ 
in Figs.~\ref{gm8h0} (b) and \ref{gm10h0} (b), respectively, which 
necessarily yields the same maximal $T_c$. 

In \fig{gm0-7h} (b), the order parameter of the $d$PI is plotted as a function of $h$ 
for a sequence of $g_m$ at low $T$. 
Because of two first order transitions at low $T$ [\fig{gm0-7h} (a)], 
the order parameter exhibits two jumps. 
Interestingly the maximal value of $\phi$ does not depend on $g_m$. 
This feature is easily understood from Eqs.~(\ref{band}) and (\ref{self-phi}). 
The right-hand side of \eq{self-phi} depends on the quantity 
$\tilde{\mu}_{\vk \sigma}=\frac{\sigma}{2}(m+h)+d_{\vk}\phi$ for a fixed $\mu$. 
Suppose the maximal value of $\phi$, say $\phi_{\rm max}$, is obtained 
at $h=h_{\rm max}$ for $g_m=0$, namely for $m=0$. 
Even when $g_m$ is turned on, the same value of $\phi_{\rm max}$ is obtained as long as 
$m$ and $h$ fulfills the equation 
\be
m+h=h_{\rm max}\,.
\label{mhmax}
\ee
This equation may hold unless the value of $m$ becomes as large as $h_{\rm max}$. 
We can check that \eq{mhmax} indeed holds up to $g_m \approx 7$, 
leading to the same maximal value of $\phi$ for $g_m=0-7$. 

In \fig{gm0-7h} (c), the magnetization is plotted as a function of $h$ at low $T$. 
Because of first order transitions at low $T$, the magnetization exhibits two successive jumps. 
It is instructive to recognize that there could occur a metamagnetic transition 
at $h\approx 0.12$ if the coupling $g_{\phi}$ would be turned off, 
indicating the underlying competition of the $d$PI and 
a metamagnetic transition. We can check that the $d$PI overcomes 
the metamagnetic transition up to $g_m=7.9$. 

\begin{figure}[ht!]
\begin{center}
\includegraphics[width=6.5cm]{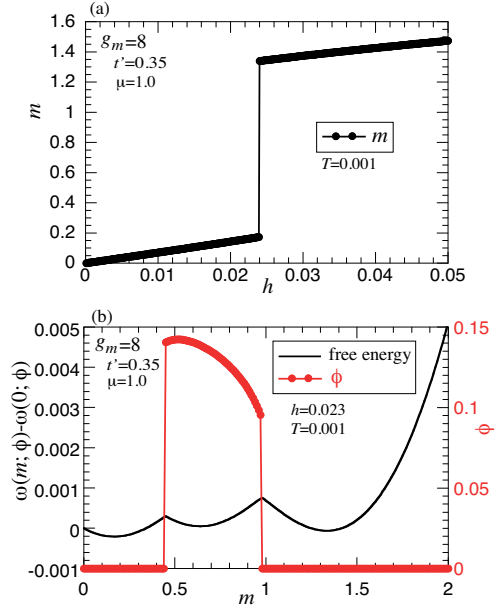}
\caption{(Color online) (a) $h$ dependence of $m$ for $g_m=8$ at $T=0.001$. 
(b) Free energy as a function of $m$ at $h=0.023$ and $T=0.001$ for $g_m=8$. 
The value of $\phi$ which minimizes the free energy at each $m$ is also plotted. 
} 
\label{gm8h}
\end{center}
\end{figure}

For $g_m \geq 8$, on the other hand, 
the metamagnetic transition becomes dominant 
and the magnetization exhibits a single jump as shown in \fig{gm8h} (a). 
The Landau free energy is plotted in \fig{gm8h} (b) as a function of $m$ at $h=0.023$, 
just below the metamagnetic transition; the order parameter $\phi$ is optimized to minimize 
the free energy at each $m$.  There are three local minima. 
Two local minima, where $\phi=0$ is stabilized, are associated with 
the metamagnetic phenomenon. 
The other local minimum, at which $\phi$ becomes finite, 
corresponds to a solution of the $d$PI. This solution, however, 
does not give the absolute minimum and thus the $d$PI does not occur.

When $g_m$ exceeds $8.25$, the FM occurs even for $h=0$. 
In this case, neither a metamagnetic transition nor a $d$PI occurs by applying a 
magnetic field.

The effect of a ferromagnetic interaction on the $d$PI for $h\ne 0$ can be summarized as follows: 
i) the $d$PI occurs in a lower magnetic field,  ii) the field range 
where the $d$PI is stabilized becomes narrower, 
iii) the first order part of the transition line extends,  
and iv) the $d$PI and a metamagnetic transition compete with each other and the former is 
realized up to $g_m \approx 8$, and the latter for $8 \lesssim g_m \lesssim 8.25$ 
for the present choice of parameters.

\section {Conclusions} 
We have studied a two-dimensional electron system, where electrons interact with each other 
via interactions favoring a $d$PI and FM. 
In the absence of a magnetic field, we have obtained rich phase diagrams. 
The $d$PI and FM typically compete with each other. In fact, while both $d$PI and FM can be 
realized simultaneously, they are separated by a first order phase boundary. 
Nevertheless it is possible that the $d$PI is stabilized inside the FM phase, leading to 
their coexistence. The presence of $t'$, leading to a breaking of 
particle-hole symmetry,  plays an important role. 
For $t'=0$, either the $d$PI or the FM is typically realized in 
the plane of the chemical potential 
and temperature, and coexistence is not stabilized. 
We have also studied the effect of a magnetic field, 
motivated by the experimental indication that Sr327 is 
in the normal state without 
a magnetic field and exhibits a nematic instability by applying a field. 
In this case, instead of FM, the $d$PI competes with a metamagnetic transition. 
The latter occurs above  a threshold strength of the FM interaction  and otherwise the $d$PI is stabilized with a dome-shaped transition line 
around the van Hove energy in the plane of a field and temperature. 
With increasing the FM interaction, 
the center of the dome shifts to a lower field, 
accompanied by a substantial reduction of the field range where the $d$PI is stabilized 
and by an extension of the first order part of the transition line, 
although the maximal $T_c$ does not change. 

It might seem that the interaction strength of $g_m$ is considered up to 
a too large value ($g_m \sim 10$) in our study. 
However, this seemingly large value is due to our definition of $g_m$ in \eq{Hm} 
where a factor of 
$(1/2)^{2}$ originating from spin is not absorbed into the definition of $g_m$. 

A typical feature of the $d$PI is that its mean-field phase diagram is 
characterized by universal ratios.\cite{yamase05,yamase07c}  
In the model solved in Ref.~\onlinecite{yamase07c}, several universal ratios 
reasonably agree with experimental values, but ratios of temperature and 
a magnetic field come out one order of magnitude smaller than the experimental data. 
For example, in experiments, 
$T_c^{\rm tri}/h_{\rm tri} \sim 0.6k_{B}/(0.15\mathfrak{g}\mu_B) = 
6\mathfrak{g}^{-1} \approx 3$ if $\mathfrak{g}=2$, whereas 
theoretically we obtain 
$T_c^{\rm tri}/h_{\rm tri}  \sim 0.3$ for $g_m=0$ (Ref.~\onlinecite{miscTCP}); 
here $h_{\rm tri}$ is the field at a tricritical point measured from the van Hove energy. 
However in the presence of a ferromagnetic interaction, 
we have found that only the scale of a magnetic field is substantially reduced 
while the temperature scale is not. 
As a result, from \fig{gm0-7h} (a),  we obtain 
$T_c^{\rm tri}/h_{\rm tri}  \sim 2$ and $7$ (Ref. \onlinecite{miscTCP})   
for $g_m=6$ and $7$, respectively.  The ratio of $T_c^{\rm tri}/h_{\rm tri}$ is 
substantially modified by a FM interaction to become 
comparable to the experimental one. 
The large value of $g_m$ indicates that the system is 
close to the FM instability for $h=0$,  
the same situation as in Sr327.\cite{ikeda00,ikeda04,capogna03,kitagawa05}

The FM interaction pushes up $T_c^{\rm tri}$ to a higher temperature,  
but the maximal $T_c$ does not change. 
As a result, other ratios such as $T_c^{\rm tri}/T_c^{\rm vH}$, where $T_c^{\rm vH}$ 
is $T_c$ at the van Hove energy, now becomes slightly larger than the experimental value, 
although it showed better agreement 
with experimental data in the model with $g_m=0$.\cite{yamase07c}  
However, this may be easily improved by invoking weak fluctuations associated with the $d$PI, 
since it was shown\cite{jakubczyk09,yamase11a} that 
fluctuations suppress $T_c^{\rm tri}$ stronger than $T_c^{\rm vH}$. 
Therefore the ratios in the experimental phase 
diagram of the $d$PI are well understood by the presence of a FM 
interaction tuning the system close to the FM instability, 
and by weak $d$PI fluctuations.  

The lines of first order phase transitions tilt outward 
in the experimental phase diagram,\cite{grigera04} 
indicating that the entropy inside the $d$PI phase is larger than that 
in the normal state.\cite{rost09} 
This counterintuitive phenomenon is not captured in the present theory. 
This inconsistency may be explored further in terms of the interplay of  
ferromagnetic fluctuations and the $d$PI by going beyond the mean-field model. 

While we have analyzed a single band model, Sr327 is a $t_{2g}$ system 
and orbital nematic order may provide another possible scenario.\cite{raghu09,wclee09} 
Since there are interactions among different orbitals, the $d$PI is expected to 
generate orbital nematic order, or vice versa. 
It is an open question which is the driving force for nematicity observed in Sr327.

\begin{acknowledgments}
The author thanks W. Metzner for 
a critical reading of the manuscript and valuable comments. 
Support by the Alexander von Humboldt Foundation 
and a Grant-in-Aid for Scientific Research from Monkasho 
is also gratefully acknowledged.  
\end{acknowledgments}

\bibliography{main.bib}

\begin{thebibliography}{67}
\expandafter\ifx\csname natexlab\endcsname\relax\def\natexlab#1{#1}\fi
\expandafter\ifx\csname bibnamefont\endcsname\relax
  \def\bibnamefont#1{#1}\fi
\expandafter\ifx\csname bibfnamefont\endcsname\relax
  \def\bibfnamefont#1{#1}\fi
\expandafter\ifx\csname citenamefont\endcsname\relax
  \def\citenamefont#1{#1}\fi
\expandafter\ifx\csname url\endcsname\relax
  \def\url#1{\texttt{#1}}\fi
\expandafter\ifx\csname urlprefix\endcsname\relax\def\urlprefix{URL }\fi
\providecommand{\bibinfo}[2]{#2}
\providecommand{\eprint}[2][]{\url{#2}}

\bibitem[{\citenamefont{de~Gennes and Prost}(1993)}]{deGennes93}
\bibinfo{author}{\bibfnamefont{P.~G.} \bibnamefont{de~Gennes}}
  \bibnamefont{and} \bibinfo{author}{\bibfnamefont{J.}~\bibnamefont{Prost}},
  \emph{\bibinfo{title}{The Physics of Liquid Crystals}}
  (\bibinfo{publisher}{Clarendon Press}, \bibinfo{address}{Oxford, UK},
  \bibinfo{year}{1993}).

\bibitem[{\citenamefont{{A. F. Andreev and I. A. Grishchuk}}(1984)}]{andreev84}
\bibinfo{author}{\bibnamefont{{A. F. Andreev and I. A. Grishchuk}}},
  \bibinfo{journal}{Sov. Phys. JETP} \textbf{\bibinfo{volume}{60}},
  \bibinfo{pages}{267} (\bibinfo{year}{1984}).

\bibitem[{\citenamefont{Misguich and Lhuillier}(2005)}]{misguich05}
\bibinfo{author}{\bibfnamefont{G.}~\bibnamefont{Misguich}} \bibnamefont{and}
  \bibinfo{author}{\bibfnamefont{C.}~\bibnamefont{Lhuillier}}, in
  \emph{\bibinfo{booktitle}{Frustrated Spin Systems}}, edited by
  \bibinfo{editor}{\bibfnamefont{H.~T.} \bibnamefont{Diep}}
  (\bibinfo{publisher}{World Scientific}, \bibinfo{address}{Singapore},
  \bibinfo{year}{2005}), p. \bibinfo{pages}{229}.

\bibitem[{\citenamefont{{S. Raghu, A. Paramekanti, E-.A. Kim, R. A. Borzi, S.
  A. Grigera, A. P. Mackenzie, and S. A. Kivelson}}(2009)}]{raghu09}
\bibinfo{author}{\bibnamefont{{S. Raghu, A. Paramekanti, E-.A. Kim, R. A.
  Borzi, S. A. Grigera, A. P. Mackenzie, and S. A. Kivelson}}},
  \bibinfo{journal}{Phys.\ Rev.\ B} \textbf{\bibinfo{volume}{79}},
  \bibinfo{pages}{214402} (\bibinfo{year}{2009}).

\bibitem[{\citenamefont{{W.-C. Lee and C. Wu}}(2009)}]{wclee09}
\bibinfo{author}{\bibnamefont{{W.-C. Lee and C. Wu}}}, \bibinfo{journal}{Phys.\
  Rev.\ B} \textbf{\bibinfo{volume}{80}}, \bibinfo{pages}{104438}
  (\bibinfo{year}{2009}).

\bibitem[{\citenamefont{{S. Kasahara, H. J. Shi, K. Hashimoto, S. Tonegawa, Y.
  Mizukami, T. Shibauchi, K. Sugimoto, T. Fukuda, T. Terashima, A. H.
  Nevidomskyy, and Y. Matsuda}}(2012)}]{kasahara12}
\bibinfo{author}{\bibnamefont{{S. Kasahara, H. J. Shi, K. Hashimoto, S.
  Tonegawa, Y. Mizukami, T. Shibauchi, K. Sugimoto, T. Fukuda, T. Terashima, A.
  H. Nevidomskyy, and Y. Matsuda}}}, \bibinfo{journal}{Nature (London)}
  \textbf{\bibinfo{volume}{486}}, \bibinfo{pages}{382} (\bibinfo{year}{2012}).

\bibitem[{\citenamefont{{S. Onari and H. Kontani}}(2012)}]{onari12}
\bibinfo{author}{\bibnamefont{{S. Onari and H. Kontani}}},
  \bibinfo{journal}{Phys. Rev. Lett.} \textbf{\bibinfo{volume}{109}},
  \bibinfo{pages}{137001} (\bibinfo{year}{2012}).

\bibitem[{\citenamefont{Kivelson et~al.}(1998)\citenamefont{Kivelson, Fradkin,
  and Emery}}]{kivelson98}
\bibinfo{author}{\bibfnamefont{S.~A.} \bibnamefont{Kivelson}},
  \bibinfo{author}{\bibfnamefont{E.}~\bibnamefont{Fradkin}}, \bibnamefont{and}
  \bibinfo{author}{\bibfnamefont{V.~J.} \bibnamefont{Emery}},
  \bibinfo{journal}{Nature (London)} \textbf{\bibinfo{volume}{393}},
  \bibinfo{pages}{550} (\bibinfo{year}{1998}).

\bibitem[{yam({\natexlab{a}})}]{yamase00}
\bibinfo{note}{{H. Yamase and H. Kohno, J.\ Phys.\ Soc.\ Jpn.\ {\bf 69}, 332
  (2000); {\bf 69}, 2151 (2000).}}

\bibitem[{\citenamefont{Halboth and Metzner}(2000)}]{metzner00}
\bibinfo{author}{\bibfnamefont{C.~J.} \bibnamefont{Halboth}} \bibnamefont{and}
  \bibinfo{author}{\bibfnamefont{W.}~\bibnamefont{Metzner}},
  \bibinfo{journal}{Phys.\ Rev.\ Lett.} \textbf{\bibinfo{volume}{85}},
  \bibinfo{pages}{5162} (\bibinfo{year}{2000}).

\bibitem[{\citenamefont{Pomeranchuk}(1959)}]{pomeranchuk59}
\bibinfo{author}{\bibfnamefont{I.~J.} \bibnamefont{Pomeranchuk}},
  \bibinfo{journal}{Sov.\ Phys.\ JETP} \textbf{\bibinfo{volume}{8}},
  \bibinfo{pages}{361} (\bibinfo{year}{1959}).

\bibitem[{\citenamefont{Miyanaga and Yamase}(2006)}]{miyanaga06}
\bibinfo{author}{\bibfnamefont{A.}~\bibnamefont{Miyanaga}} \bibnamefont{and}
  \bibinfo{author}{\bibfnamefont{H.}~\bibnamefont{Yamase}},
  \bibinfo{journal}{Phys. Rev. B} \textbf{\bibinfo{volume}{73}},
  \bibinfo{pages}{174513} (\bibinfo{year}{2006}).

\bibitem[{\citenamefont{{B. Edegger, V. N. Muthukumar, and C.
  Gros}}(2006)}]{edegger06}
\bibinfo{author}{\bibnamefont{{B. Edegger, V. N. Muthukumar, and C. Gros}}},
  \bibinfo{journal}{Phys.\ Rev.\ B} \textbf{\bibinfo{volume}{74}},
  \bibinfo{pages}{165109} (\bibinfo{year}{2006}).

\bibitem[{\citenamefont{{M. Bejas, A. Greco, and H. Yamase}}(2012)}]{bejas12}
\bibinfo{author}{\bibnamefont{{M. Bejas, A. Greco, and H. Yamase}}},
  \bibinfo{journal}{Phys.\ Rev.\ B} \textbf{\bibinfo{volume}{86}},
  \bibinfo{pages}{224509} (\bibinfo{year}{2012}).

\bibitem[{weg()}]{wegner02}
\bibinfo{note}{I. Grote, E. K{\"{o}}rding, and F. Wegner, J.\ Low\ Temp.\
  Phys.\ {\bf 126}, 1385 (2002); V. Hankevych, I. Grote, and F. Wegner, Phys.\
  Rev.\ B \ {\bf 66}, 094516 (2002)}.

\bibitem[{\citenamefont{Neumayr and Metzner}(2003)}]{neumayr03}
\bibinfo{author}{\bibfnamefont{A.}~\bibnamefont{Neumayr}} \bibnamefont{and}
  \bibinfo{author}{\bibfnamefont{W.}~\bibnamefont{Metzner}},
  \bibinfo{journal}{Phys.\ Rev.\ B} \textbf{\bibinfo{volume}{67}},
  \bibinfo{pages}{035112} (\bibinfo{year}{2003}).

\bibitem[{\citenamefont{{S. Okamoto, D. S\'{e}n\'{e}chal, M. Civelli, and A.-M.
  Tremblay}}(2010)}]{okamoto10}
\bibinfo{author}{\bibnamefont{{S. Okamoto, D. S\'{e}n\'{e}chal, M. Civelli, and
  A.-M. Tremblay}}}, \bibinfo{journal}{Phys.\ Rev.\ B}
  \textbf{\bibinfo{volume}{82}}, \bibinfo{pages}{180511}
  (\bibinfo{year}{2010}).

\bibitem[{\citenamefont{{S.-Q. Su and T. A. Maier}}(2011)}]{su11}
\bibinfo{author}{\bibnamefont{{S.-Q. Su and T. A. Maier}}},
  \bibinfo{journal}{Phys.\ Rev.\ B} \textbf{\bibinfo{volume}{84}},
  \bibinfo{pages}{220506(R)} (\bibinfo{year}{2011}).

\bibitem[{\citenamefont{{J. Buenemann, T. Schickling, and F.
  Gebhard}}(2012)}]{buenemann12}
\bibinfo{author}{\bibnamefont{{J. Buenemann, T. Schickling, and F. Gebhard}}},
  \bibinfo{journal}{Europhys. Lett.} \textbf{\bibinfo{volume}{98}},
  \bibinfo{pages}{27006} (\bibinfo{year}{2012}).

\bibitem[{\citenamefont{Khavkine et~al.}(2004)\citenamefont{Khavkine, Chung,
  Oganesyan, and Kee}}]{khavkine04}
\bibinfo{author}{\bibfnamefont{I.}~\bibnamefont{Khavkine}},
  \bibinfo{author}{\bibfnamefont{C.-H.} \bibnamefont{Chung}},
  \bibinfo{author}{\bibfnamefont{V.}~\bibnamefont{Oganesyan}},
  \bibnamefont{and} \bibinfo{author}{\bibfnamefont{H.-Y.} \bibnamefont{Kee}},
  \bibinfo{journal}{Phys.\ Rev.\ B} \textbf{\bibinfo{volume}{70}},
  \bibinfo{pages}{155110} (\bibinfo{year}{2004}).

\bibitem[{\citenamefont{Yamase et~al.}(2005)\citenamefont{Yamase, Oganesyan,
  and Metzner}}]{yamase05}
\bibinfo{author}{\bibfnamefont{H.}~\bibnamefont{Yamase}},
  \bibinfo{author}{\bibfnamefont{V.}~\bibnamefont{Oganesyan}},
  \bibnamefont{and} \bibinfo{author}{\bibfnamefont{W.}~\bibnamefont{Metzner}},
  \bibinfo{journal}{Phys.\ Rev.\ B} \textbf{\bibinfo{volume}{72}},
  \bibinfo{pages}{035114} (\bibinfo{year}{2005}).

\bibitem[{\citenamefont{Quintanilla and Schofield}(2006)}]{quintanilla06}
\bibinfo{author}{\bibfnamefont{J.}~\bibnamefont{Quintanilla}} \bibnamefont{and}
  \bibinfo{author}{\bibfnamefont{A.~J.} \bibnamefont{Schofield}},
  \bibinfo{journal}{Phys.\ Rev.\ B} \textbf{\bibinfo{volume}{74}},
  \bibinfo{pages}{115126} (\bibinfo{year}{2006}).

\bibitem[{\citenamefont{{J. Quintanilla, M. Haque, and A. J.
  Schofield}}(2008)}]{quintanilla08}
\bibinfo{author}{\bibnamefont{{J. Quintanilla, M. Haque, and A. J.
  Schofield}}}, \bibinfo{journal}{Phys.\ Rev.\ B}
  \textbf{\bibinfo{volume}{78}}, \bibinfo{pages}{035131}
  (\bibinfo{year}{2008}).

\bibitem[{\citenamefont{{C. A. Lamas, D. C. Cabra, and N.
  Grandi}}(2008)}]{lamas08}
\bibinfo{author}{\bibnamefont{{C. A. Lamas, D. C. Cabra, and N. Grandi}}},
  \bibinfo{journal}{Phys.\ Rev.\ B} \textbf{\bibinfo{volume}{78}},
  \bibinfo{pages}{115104} (\bibinfo{year}{2008}).

\bibitem[{\citenamefont{{M. V. Zverev, J. W. Clark, Z. Nussinov, and V. A.
  Khodel}}(2010)}]{zverev10}
\bibinfo{author}{\bibnamefont{{M. V. Zverev, J. W. Clark, Z. Nussinov, and V.
  A. Khodel}}}, \bibinfo{journal}{Phys.\ Rev.\ B}
  \textbf{\bibinfo{volume}{82}}, \bibinfo{pages}{125111}
  (\bibinfo{year}{2010}).

\bibitem[{\citenamefont{Oganesyan et~al.}(2001)\citenamefont{Oganesyan,
  Kivelson, and Fradkin}}]{oganesyan01}
\bibinfo{author}{\bibfnamefont{V.}~\bibnamefont{Oganesyan}},
  \bibinfo{author}{\bibfnamefont{S.~A.} \bibnamefont{Kivelson}},
  \bibnamefont{and} \bibinfo{author}{\bibfnamefont{E.}~\bibnamefont{Fradkin}},
  \bibinfo{journal}{Phys.\ Rev.\ B} \textbf{\bibinfo{volume}{64}},
  \bibinfo{pages}{195109} (\bibinfo{year}{2001}).

\bibitem[{\citenamefont{{D. Barci and L. E. Oxman}}(2003)}]{barci03}
\bibinfo{author}{\bibnamefont{{D. Barci and L. E. Oxman}}},
  \bibinfo{journal}{Phys.\ Rev.\ B} \textbf{\bibinfo{volume}{67}},
  \bibinfo{pages}{205108} (\bibinfo{year}{2003}).

\bibitem[{\citenamefont{{J. Nilsson and A. H. Castro Neto}}(2005)}]{nilsson05}
\bibinfo{author}{\bibnamefont{{J. Nilsson and A. H. Castro Neto}}},
  \bibinfo{journal}{Phys.\ Rev.\ B} \textbf{\bibinfo{volume}{72}},
  \bibinfo{pages}{195104} (\bibinfo{year}{2005}).

\bibitem[{\citenamefont{{M. J. Lawler, D. G. Barci, V. Fern{\'a}ndez, E.
  Fradkin, and L. Oxman}}(2006)}]{lawler06}
\bibinfo{author}{\bibnamefont{{M. J. Lawler, D. G. Barci, V. Fern{\'a}ndez, E.
  Fradkin, and L. Oxman}}}, \bibinfo{journal}{Phys.\ Rev.\ B}
  \textbf{\bibinfo{volume}{73}}, \bibinfo{pages}{085101}
  (\bibinfo{year}{2006}).

\bibitem[{\citenamefont{{M. Zacharias, P. W{\"o}lfle, and M.
  Garst}}(2009)}]{zacharias09}
\bibinfo{author}{\bibnamefont{{M. Zacharias, P. W{\"o}lfle, and M. Garst}}},
  \bibinfo{journal}{Phys.\ Rev.\ B} \textbf{\bibinfo{volume}{80}},
  \bibinfo{pages}{165116} (\bibinfo{year}{2009}).

\bibitem[{\citenamefont{{D. L. Maslov and A. V. Chubukov}}(2010)}]{maslov10}
\bibinfo{author}{\bibnamefont{{D. L. Maslov and A. V. Chubukov}}},
  \bibinfo{journal}{Phys.\ Rev.\ B} \textbf{\bibinfo{volume}{81}},
  \bibinfo{pages}{045110} (\bibinfo{year}{2010}).

\bibitem[{\citenamefont{{P. Jakubczyk, W. Metzner, and H.
  Yamase}}(2009)}]{jakubczyk09}
\bibinfo{author}{\bibnamefont{{P. Jakubczyk, W. Metzner, and H. Yamase}}},
  \bibinfo{journal}{Phys. Rev. Lett.} \textbf{\bibinfo{volume}{103}},
  \bibinfo{pages}{220602} (\bibinfo{year}{2009}).

\bibitem[{\citenamefont{{H. Yamase, P. Jakubczyk, and W.
  Metzner}}(2011)}]{yamase11a}
\bibinfo{author}{\bibnamefont{{H. Yamase, P. Jakubczyk, and W. Metzner}}},
  \bibinfo{journal}{Phys. Rev. B} \textbf{\bibinfo{volume}{83}},
  \bibinfo{pages}{125121} (\bibinfo{year}{2011}).

\bibitem[{\citenamefont{Metzner et~al.}(2003)\citenamefont{Metzner, Rohe, and
  Andergassen}}]{metzner03}
\bibinfo{author}{\bibfnamefont{W.}~\bibnamefont{Metzner}},
  \bibinfo{author}{\bibfnamefont{D.}~\bibnamefont{Rohe}}, \bibnamefont{and}
  \bibinfo{author}{\bibfnamefont{S.}~\bibnamefont{Andergassen}},
  \bibinfo{journal}{Phys.\ Rev.\ Lett.} \textbf{\bibinfo{volume}{91}},
  \bibinfo{pages}{066402} (\bibinfo{year}{2003}).

\bibitem[{\citenamefont{Dell'Anna and Metzner}(2006)}]{dellanna06}
\bibinfo{author}{\bibfnamefont{L.}~\bibnamefont{Dell'Anna}} \bibnamefont{and}
  \bibinfo{author}{\bibfnamefont{W.}~\bibnamefont{Metzner}},
  \bibinfo{journal}{Phys.\ Rev.\ B} \textbf{\bibinfo{volume}{73}},
  \bibinfo{pages}{045127} (\bibinfo{year}{2006}).

\bibitem[{\citenamefont{Yamase and Metzner}(2012)}]{yamase12}
\bibinfo{author}{\bibfnamefont{H.}~\bibnamefont{Yamase}} \bibnamefont{and}
  \bibinfo{author}{\bibfnamefont{W.}~\bibnamefont{Metzner}},
  \bibinfo{journal}{Phys. Rev. Lett.} \textbf{\bibinfo{volume}{108}},
  \bibinfo{pages}{186405} (\bibinfo{year}{2012}).

\bibitem[{\citenamefont{Hinkov et~al.}(2004)\citenamefont{Hinkov, Pailh\`{e}s,
  Bourges, Sidis, Ivanov, Kulakov, Lin, Chen, Bernhard, and Keimer}}]{hinkov04}
\bibinfo{author}{\bibfnamefont{V.}~\bibnamefont{Hinkov}},
  \bibinfo{author}{\bibfnamefont{S.}~\bibnamefont{Pailh\`{e}s}},
  \bibinfo{author}{\bibfnamefont{P.}~\bibnamefont{Bourges}},
  \bibinfo{author}{\bibfnamefont{Y.}~\bibnamefont{Sidis}},
  \bibinfo{author}{\bibfnamefont{A.}~\bibnamefont{Ivanov}},
  \bibinfo{author}{\bibfnamefont{A.}~\bibnamefont{Kulakov}},
  \bibinfo{author}{\bibfnamefont{C.~T.} \bibnamefont{Lin}},
  \bibinfo{author}{\bibfnamefont{D.}~\bibnamefont{Chen}},
  \bibinfo{author}{\bibfnamefont{C.}~\bibnamefont{Bernhard}}, \bibnamefont{and}
  \bibinfo{author}{\bibfnamefont{B.}~\bibnamefont{Keimer}},
  \bibinfo{journal}{Nature (London)} \textbf{\bibinfo{volume}{430}},
  \bibinfo{pages}{650} (\bibinfo{year}{2004}).

\bibitem[{\citenamefont{{V. Hinkov, P. Bourges, S. Pailh\`{e}s, Y. Sidis, A.
  Ivanov, C. D. Frost, T. G. Perring, C. T. Lin, D. P. Chen, and B.
  Keimer}}(2007)}]{hinkov07}
\bibinfo{author}{\bibnamefont{{V. Hinkov, P. Bourges, S. Pailh\`{e}s, Y. Sidis,
  A. Ivanov, C. D. Frost, T. G. Perring, C. T. Lin, D. P. Chen, and B.
  Keimer}}}, \bibinfo{journal}{Nat. Phys.} \textbf{\bibinfo{volume}{3}},
  \bibinfo{pages}{780} (\bibinfo{year}{2007}).

\bibitem[{\citenamefont{{V. Hinkov, D. Haug, B. Fauqu\'{e}, P. Bourges, Y.
  Sidis, A. Ivanov, C. Bernhard, C. T. Lin, and B. Keimer}}(2008)}]{hinkov08}
\bibinfo{author}{\bibnamefont{{V. Hinkov, D. Haug, B. Fauqu\'{e}, P. Bourges,
  Y. Sidis, A. Ivanov, C. Bernhard, C. T. Lin, and B. Keimer}}},
  \bibinfo{journal}{Science} \textbf{\bibinfo{volume}{319}},
  \bibinfo{pages}{597} (\bibinfo{year}{2008}).

\bibitem[{\citenamefont{Yamase and Metzner}(2006)}]{yamase06}
\bibinfo{author}{\bibfnamefont{H.}~\bibnamefont{Yamase}} \bibnamefont{and}
  \bibinfo{author}{\bibfnamefont{W.}~\bibnamefont{Metzner}},
  \bibinfo{journal}{Phys.\ Rev.\ B} \textbf{\bibinfo{volume}{73}},
  \bibinfo{pages}{214517} (\bibinfo{year}{2006}).

\bibitem[{\citenamefont{Yamase}(2009)}]{yamase09}
\bibinfo{author}{\bibfnamefont{H.}~\bibnamefont{Yamase}},
  \bibinfo{journal}{Phys.\ Rev.\ B} \textbf{\bibinfo{volume}{79}},
  \bibinfo{pages}{052501} (\bibinfo{year}{2009}).

\bibitem[{\citenamefont{{R. Daou, J. Chang, D. LeBoeuf, O. Cyr-Choini\`{e}re,
  F. Lalibert\'{e}, N. Doiron-Leyraud, B. J. Ramshaw, R. Liang, D. A. Bonn, W.
  H. Hardy, and L. Taillefer}}(2010)}]{daou10}
\bibinfo{author}{\bibnamefont{{R. Daou, J. Chang, D. LeBoeuf, O.
  Cyr-Choini\`{e}re, F. Lalibert\'{e}, N. Doiron-Leyraud, B. J. Ramshaw, R.
  Liang, D. A. Bonn, W. H. Hardy, and L. Taillefer}}}, \bibinfo{journal}{Nature
  (London)} \textbf{\bibinfo{volume}{463}}, \bibinfo{pages}{519}
  (\bibinfo{year}{2010}).

\bibitem[{\citenamefont{Hackl and Vojta}(2009)}]{hackl09}
\bibinfo{author}{\bibfnamefont{A.}~\bibnamefont{Hackl}} \bibnamefont{and}
  \bibinfo{author}{\bibfnamefont{M.}~\bibnamefont{Vojta}},
  \bibinfo{journal}{Phys.\ Rev.\ B} \textbf{\bibinfo{volume}{80}},
  \bibinfo{pages}{220514(R)} (\bibinfo{year}{2009}).

\bibitem[{\citenamefont{{S. A. Grigera, P. Gegenwart, R. A. Borzi, F. Weickert,
  A. J. Schofield, R. S. Perry, T. Tayama, T. Sakakibara, Y. Maeno, A. G.
  Green, and A. P. Mackenzie}}(2004)}]{grigera04}
\bibinfo{author}{\bibnamefont{{S. A. Grigera, P. Gegenwart, R. A. Borzi, F.
  Weickert, A. J. Schofield, R. S. Perry, T. Tayama, T. Sakakibara, Y. Maeno,
  A. G. Green, and A. P. Mackenzie}}}, \bibinfo{journal}{Science}
  \textbf{\bibinfo{volume}{306}}, \bibinfo{pages}{1154} (\bibinfo{year}{2004}).

\bibitem[{\citenamefont{{R. A. Borzi, S. A. Grigera, J. Farrell, R. S. Perry,
  S. J. S. Lister, S. L. Lee, D. A. Tennant, Y. Maeno, and A. P.
  Mackenzie}}(2007)}]{borzi07}
\bibinfo{author}{\bibnamefont{{R. A. Borzi, S. A. Grigera, J. Farrell, R. S.
  Perry, S. J. S. Lister, S. L. Lee, D. A. Tennant, Y. Maeno, and A. P.
  Mackenzie}}}, \bibinfo{journal}{Science} \textbf{\bibinfo{volume}{315}},
  \bibinfo{pages}{214} (\bibinfo{year}{2007}).

\bibitem[{\citenamefont{{A. W. Rost, R. S. Perry, J.-F. Mercure, A. P.
  Mackenzie, and S. A. Grigera}}(2009)}]{rost09}
\bibinfo{author}{\bibnamefont{{A. W. Rost, R. S. Perry, J.-F. Mercure, A. P.
  Mackenzie, and S. A. Grigera}}}, \bibinfo{journal}{Science}
  \textbf{\bibinfo{volume}{325}}, \bibinfo{pages}{1360} (\bibinfo{year}{2009}).

\bibitem[{\citenamefont{Kee and Kim}(2005)}]{kee05}
\bibinfo{author}{\bibfnamefont{H.-Y.} \bibnamefont{Kee}} \bibnamefont{and}
  \bibinfo{author}{\bibfnamefont{Y.~B.} \bibnamefont{Kim}},
  \bibinfo{journal}{Phys.\ Rev.\ B} \textbf{\bibinfo{volume}{71}},
  \bibinfo{pages}{184402} (\bibinfo{year}{2005}).

\bibitem[{\citenamefont{Doh et~al.}(2007)\citenamefont{Doh, Kim, and
  Ahn}}]{doh07}
\bibinfo{author}{\bibfnamefont{H.}~\bibnamefont{Doh}},
  \bibinfo{author}{\bibfnamefont{Y.~B.} \bibnamefont{Kim}}, \bibnamefont{and}
  \bibinfo{author}{\bibfnamefont{K.~H.} \bibnamefont{Ahn}},
  \bibinfo{journal}{Phys.\ Rev.\ Lett.} \textbf{\bibinfo{volume}{98}},
  \bibinfo{pages}{126407} (\bibinfo{year}{2007}).

\bibitem[{\citenamefont{{C. Puetter, H. Doh, and H.-Y. Kee}}(2007)}]{puetter07}
\bibinfo{author}{\bibnamefont{{C. Puetter, H. Doh, and H.-Y. Kee}}},
  \bibinfo{journal}{Phys.\ Rev.\ B} \textbf{\bibinfo{volume}{76}},
  \bibinfo{pages}{235112} (\bibinfo{year}{2007}).

\bibitem[{yam({\natexlab{b}})}]{yamase09ab}
\bibinfo{note}{H. Yamase, Phys. Rev. Lett. {\bf 102}, 116404 (2009); Phys. Rev.
  B {\bf 80}, 115102 (2009).}

\bibitem[{\citenamefont{Ho and Schofield}(2008)}]{ho08}
\bibinfo{author}{\bibfnamefont{A.~F.} \bibnamefont{Ho}} \bibnamefont{and}
  \bibinfo{author}{\bibfnamefont{A.~J.} \bibnamefont{Schofield}},
  \bibinfo{journal}{Europhys. Lett.} \textbf{\bibinfo{volume}{84}},
  \bibinfo{pages}{27007} (\bibinfo{year}{2008}).

\bibitem[{\citenamefont{{M. H. Fischer, and M. Sigrist}}(2010)}]{fischer10}
\bibinfo{author}{\bibnamefont{{M. H. Fischer, and M. Sigrist}}},
  \bibinfo{journal}{Phys. Rev. B} \textbf{\bibinfo{volume}{81}},
  \bibinfo{pages}{064435} (\bibinfo{year}{2010}).

\bibitem[{yam({\natexlab{c}})}]{yamase0710}
\bibinfo{note}{H. Yamase and A. A. Katanin, J.\ Phys.\ Soc.\ Jpn.\ {\bf 76},
  073706 (2007); {\bf 79}, 127001 (2010).}

\bibitem[{\citenamefont{Yamase}(2007)}]{yamase07c}
\bibinfo{author}{\bibfnamefont{H.}~\bibnamefont{Yamase}},
  \bibinfo{journal}{Phys.\ Rev.\ B} \textbf{\bibinfo{volume}{76}},
  \bibinfo{pages}{155117} (\bibinfo{year}{2007}).

\bibitem[{\citenamefont{{S.-I. Ikeda, Y. Maeno, S. Nakatsuji, M. Kosaka, and Y.
  Uwatoko}}(2000)}]{ikeda00}
\bibinfo{author}{\bibnamefont{{S.-I. Ikeda, Y. Maeno, S. Nakatsuji, M. Kosaka,
  and Y. Uwatoko}}}, \bibinfo{journal}{Phys.\ Rev.\ B}
  \textbf{\bibinfo{volume}{62}}, \bibinfo{pages}{R6089} (\bibinfo{year}{2000}).

\bibitem[{\citenamefont{{S.-I. Ikeda, N. Shirakawa, T. Yanagisawa, Y. Yoshida,
  S. Koikegami, S. Koike, M. Kosaka, and Y. Uwatoko}}(2004)}]{ikeda04}
\bibinfo{author}{\bibnamefont{{S.-I. Ikeda, N. Shirakawa, T. Yanagisawa, Y.
  Yoshida, S. Koikegami, S. Koike, M. Kosaka, and Y. Uwatoko}}},
  \bibinfo{journal}{J.\ Phys.\ Soc.\ Jpn.} \textbf{\bibinfo{volume}{73}},
  \bibinfo{pages}{1322} (\bibinfo{year}{2004}).

\bibitem[{\citenamefont{{L. Capogna, E. M. Forgan, S. M. Hayden, A. Wildes, J.
  A. Duffy, A. P. Mackenzie, R. S. Perry, S. Ikeda, Y. Maeno, and S. P.
  Brown}}(2003)}]{capogna03}
\bibinfo{author}{\bibnamefont{{L. Capogna, E. M. Forgan, S. M. Hayden, A.
  Wildes, J. A. Duffy, A. P. Mackenzie, R. S. Perry, S. Ikeda, Y. Maeno, and S.
  P. Brown}}}, \bibinfo{journal}{Phys.\ Rev.\ B} \textbf{\bibinfo{volume}{67}},
  \bibinfo{pages}{012504} (\bibinfo{year}{2003}).

\bibitem[{\citenamefont{{K. Kitagawa, K. Ishida, R. S. Perry, T. Tayama, T.
  Sakakibara, and Y. Maeno}}(2005)}]{kitagawa05}
\bibinfo{author}{\bibnamefont{{K. Kitagawa, K. Ishida, R. S. Perry, T. Tayama,
  T. Sakakibara, and Y. Maeno}}}, \bibinfo{journal}{Phys.\ Rev.\ Lett.}
  \textbf{\bibinfo{volume}{95}}, \bibinfo{pages}{127001}
  (\bibinfo{year}{2005}).

\bibitem[{\citenamefont{{P. Gegenwart, F. Weickert, M. Garst, R. S. Perry, and
  Y. Maeno}}(2006)}]{gegenwart06}
\bibinfo{author}{\bibnamefont{{P. Gegenwart, F. Weickert, M. Garst, R. S.
  Perry, and Y. Maeno}}}, \bibinfo{journal}{Phys.\ Rev.\ Lett.}
  \textbf{\bibinfo{volume}{96}}, \bibinfo{pages}{136402}
  (\bibinfo{year}{2006}).

\bibitem[{\citenamefont{Hase and Nishihara}(1997)}]{hase97}
\bibinfo{author}{\bibfnamefont{I.}~\bibnamefont{Hase}} \bibnamefont{and}
  \bibinfo{author}{\bibfnamefont{Y.}~\bibnamefont{Nishihara}},
  \bibinfo{journal}{J. Phys. Soc. Jpn.} \textbf{\bibinfo{volume}{66}},
  \bibinfo{pages}{3517} (\bibinfo{year}{1997}).

\bibitem[{\citenamefont{Singh and Mazin}(2001)}]{singh01}
\bibinfo{author}{\bibfnamefont{D.~J.} \bibnamefont{Singh}} \bibnamefont{and}
  \bibinfo{author}{\bibfnamefont{I.~I.} \bibnamefont{Mazin}},
  \bibinfo{journal}{Phys.\ Rev.\ B} \textbf{\bibinfo{volume}{63}},
  \bibinfo{pages}{165101} (\bibinfo{year}{2001}).

\bibitem[{\citenamefont{{A. J. Millis, A. J. Schofield, G. G. Lonzarich, and S.
  A. Grigera}}(2002)}]{millis02}
\bibinfo{author}{\bibnamefont{{A. J. Millis, A. J. Schofield, G. G. Lonzarich,
  and S. A. Grigera}}}, \bibinfo{journal}{Phys.\ Rev.\ Lett.}
  \textbf{\bibinfo{volume}{88}}, \bibinfo{pages}{217204}
  (\bibinfo{year}{2002}).

\bibitem[{\citenamefont{{B. Binz and M. Sigrist}}(2004)}]{binz04}
\bibinfo{author}{\bibnamefont{{B. Binz and M. Sigrist}}},
  \bibinfo{journal}{Europhys. Lett.} \textbf{\bibinfo{volume}{65}},
  \bibinfo{pages}{816} (\bibinfo{year}{2004}).

\bibitem[{\citenamefont{{R. S. Perry, L. M. Galvin, S. A. Grigera, L. Capogna,
  A. J. Schofield, A. P. Mackenzie, M. Chiao, S. R. Julian, S. I. Ikeda, S.
  Nakatsuji, Y. Maeno, and C. Pfleiderer}}(2001)}]{perry01}
\bibinfo{author}{\bibnamefont{{R. S. Perry, L. M. Galvin, S. A. Grigera, L.
  Capogna, A. J. Schofield, A. P. Mackenzie, M. Chiao, S. R. Julian, S. I.
  Ikeda, S. Nakatsuji, Y. Maeno, and C. Pfleiderer}}}, \bibinfo{journal}{Phys.\
  Rev.\ Lett.} \textbf{\bibinfo{volume}{86}}, \bibinfo{pages}{2661}
  (\bibinfo{year}{2001}).

\bibitem[{\citenamefont{Valenzuela and Vozmediano}(2001)}]{valenzuela01}
\bibinfo{author}{\bibfnamefont{B.}~\bibnamefont{Valenzuela}} \bibnamefont{and}
  \bibinfo{author}{\bibfnamefont{M.~A.~H.} \bibnamefont{Vozmediano}},
  \bibinfo{journal}{Phys.\ Rev.\ B} \textbf{\bibinfo{volume}{63}},
  \bibinfo{pages}{153103} (\bibinfo{year}{2001}).

\bibitem[{\citenamefont{{M. H. Fischer, and E-.A. Kim}}(2011)}]{fischer11}
\bibinfo{author}{\bibnamefont{{M. H. Fischer, and E-.A. Kim}}},
  \bibinfo{journal}{Phys. Rev. B} \textbf{\bibinfo{volume}{84}},
  \bibinfo{pages}{144502} (\bibinfo{year}{2011}).

\bibitem[{mis()}]{miscTCP}
\bibinfo{note}{We have taken an average of two different ratios of $T_c^{\rm
  tri}/h_{\rm tri}$ because of the presence of two tricritical points.}

\end{thebibliography}

\end{document}